\documentclass[a4paper,11pt]{article}
\pdfoutput=1 
\usepackage{jinstpub}
\usepackage{lineno}
\usepackage{stackengine}
\usepackage{graphicx}
\usepackage{overpic}
\usepackage{tikz}    

\title{\boldmath \center A new Machine Learning-based method for identification of time-correlated events at tagged photon facilities}

\author[a,1]{V. Sokhoyan,}\note{Corresponding author.}
\author[a]{E. Mornacchi}

\affiliation[a]{Institut f\"ur Kernphysik, University of Mainz, Johann-Joachim-Becher-Weg 45, D-55099 Mainz, Germany}
\emailAdd{sokhoyan@uni-mainz.de}

\abstract{We present a new Machine Learning-based multivariate  analysis method for the selection of time-correlated hits in the tagging system and devices used to detect particles in the final state at the bremsstrahlung-based tagged photon facilities. This method can be applied instead of the widely used sampling and subtraction of the time-uncorrelated background, in particular at experiments aiming for high precision, where the subtraction of the time-uncorrelated background leads to increased uncertainties. Moreover, the identification of events with Machine Learning algorithms allows to preserve the information about correlations of kinematic variables in the final state, which can be advantageous for further phenomenological analyses of the experimental results.} 


\begin{document}
\maketitle
\flushbottom

\section{Introduction}
\label{sec:intro}

The accuracy of the measurements performed at electron accelerators with bremsstrahlung-based tagged photons is limited, among other factors, by the presence of the time-uncorrelated background in the tagging system at high rates, complicating unambiguous matching of the initial electron and the bremsstrahlung photon inducing the reaction in the target. The widely used sampling and subtraction of time-correlated events leads to a reduction in the measurement accuracy, depending on the degree of contamination of the data sample with time-uncorrelated background.
The method presented in this paper allows for the identification of the time-correlated events without subtraction of the random (uncorrelated) background.
To illustrate the application of the method, we have chosen the reaction $\gamma p \to p \pi^{0}$ for a selected photon beam energy range of $240-260$~MeV, where no other reactions are expected to contribute to the sample with two photons and one proton candidate in the final state.
In addition, kinematic cuts were applied for reaction identification and background suppression (see section~\ref{subsec: input}).
After the initial event selection, the time-correlated (signal) events in the experimental data are expected to have the properties of a simulated $\gamma p \to p \pi^{0}$ reaction, under the assumption that the simulation reasonably describes the data.
Afterwards, the samples obtained from Monte Carlo simulation of the signal and experimental measurement of the time-uncorrelated (random) background are used to create Machine Learning (ML) models, which are applied to distinguish between time-correlated and uncorrelated events in the experimental data.

This paper is organized as follows. Section~\ref{sec:setup} introduces the experimental setup and the conventional method for the subtraction of the random background coincidences in the tagging system.
Section~\ref{label: concept} explains the concept of the new method, while the application of the ML models, including the comparison of the new method with the conventional subtraction approach, is discussed in section~\ref{label: method}.
The outcome of this work is summarized in section~\ref{label: Summary}.

\section{Selection of time-correlated hits with a conventional approach}
\label{sec:setup}

\begin{figure}[hbtp]
\centering 
\includegraphics[width=\textwidth]{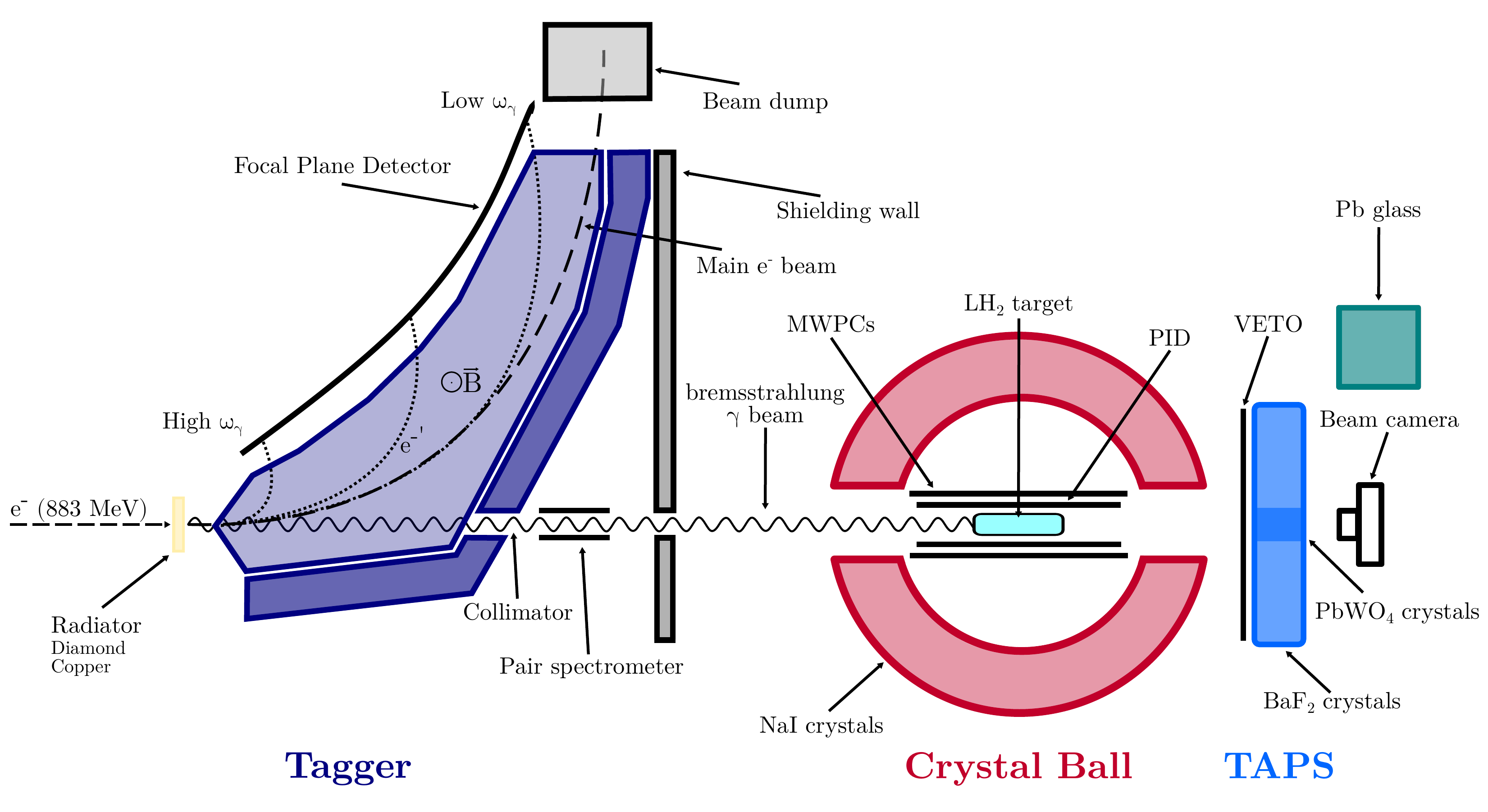}
\caption{\label{fig:setup} Schematic setup of the Crystal Ball/TAPS experiment at MAMI~\cite{bib:Mornacchi:2021}.}
\end{figure}

In this paper, the application of a new method for the identification of time-correlated events with ML-based models is illustrated for experimental data obtained with the Crystal Ball/TAPS setup at the Mainz Microtron (MAMI)~\cite{bib:Jankowiak:2006,bib:Kaiser:2008}. Figure~\ref{fig:setup} shows a schematic picture of the Crystal Ball/TAPS experiment. The photon beam is produced via bremsstrahlung by the electron beam from MAMI on a thin radiator. The outgoing electrons are bent by a dipole magnet and detected by the tagger spectrometer~\cite{bib:McGeorge:2007}, while the remaining part of the electron beam is directed to the electron beam dump. The energy of the bremsstrahlung photons is determined as the difference between the energy of the electron beam and the energy of the deflected electrons, measured by the tagger. The current setup of the tagger includes 408 channels, divided into 51 modules. Each channel is composed of a plastic scintillator (EJ200) rod, $30$~mm long with a $6 \times 6$~mm base, read out by a $6 \times 6$~mm SensL-SiPM with a bias voltage of $25$~mV. The signal is then guided outside the region with intense radiation by long Ethernet cables and is fed to a Constant Fraction Discriminator (CFD). This configuration ensures a single-counter time resolution of $\delta t = 0.1$~ns. The bremsstrahlung photons can be tagged at $4.3\%$ to $93.0\%$ of the incoming electron beam energy $E_{e}$. The energy resolution relative to $E_{e}$ varies over the energy spectrum, from low to high photon energies, from $0.4\%$ to $0.11\%$, and the absolute energy resolution varies from $3.47$~MeV to $1.03$~MeV respectively~\cite{bib:Mornacchi:2021}. The resulting photon beam (after collimation) impinges on a $10$~cm-long LH$_2$ target and the particles in the final state are detected by the nearly $4\pi$ Crystal Ball/TAPS detector system consisting of the Crystal Ball (CB)~\cite{bib:Nefkens:1995} and TAPS~\cite{bib:Gabler:1994,bib:Novotny:1998} calorimeters (and other charged particle detectors), covering $\approx 97\%$ of the solid angle.
Additional information on the apparatus can be found in refs.~\cite{bib:Unverzagt:2008,bib:Mornacchi:2021}. 
Due to the relatively high electron beam current, the number of hits in the tagger associated with the same event in the CB/TAPS setup can reach large numbers (typically up to $\approx 140$ electrons in the same trigger window), making it difficult to correlate the event with the correct hit in the tagger.
In a conventional approach, this is achieved by calculating the time difference $\Delta t$ between each hit in the tagger and the event in the CB/TAPS system.
This allows for the selection of the time-coincident events with a subsequent sampling and subtraction of the remaining time-uncorrelated background. 

\begin{figure}[htbp]
\centering 
\includegraphics[width=0.49\textwidth]{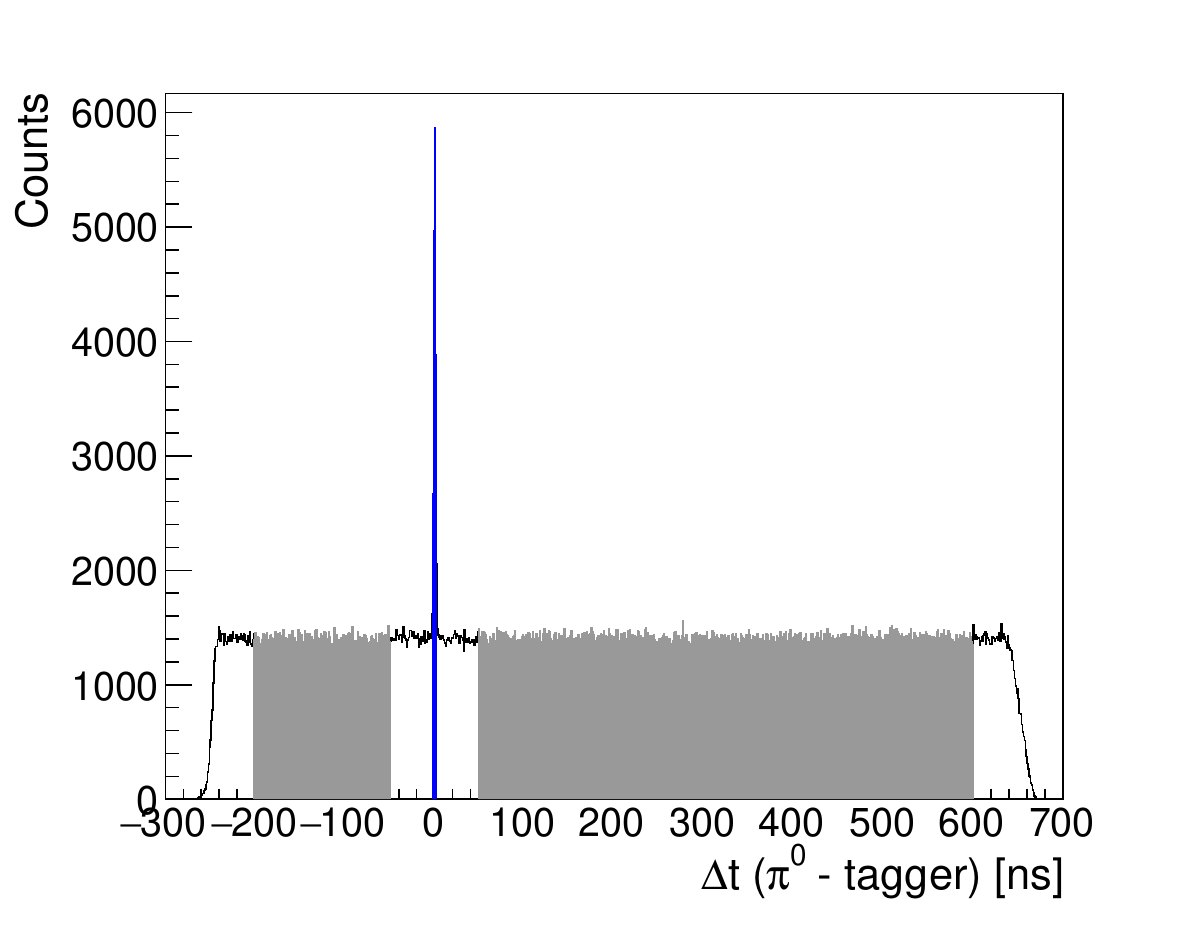}
\includegraphics[width=0.49\textwidth]{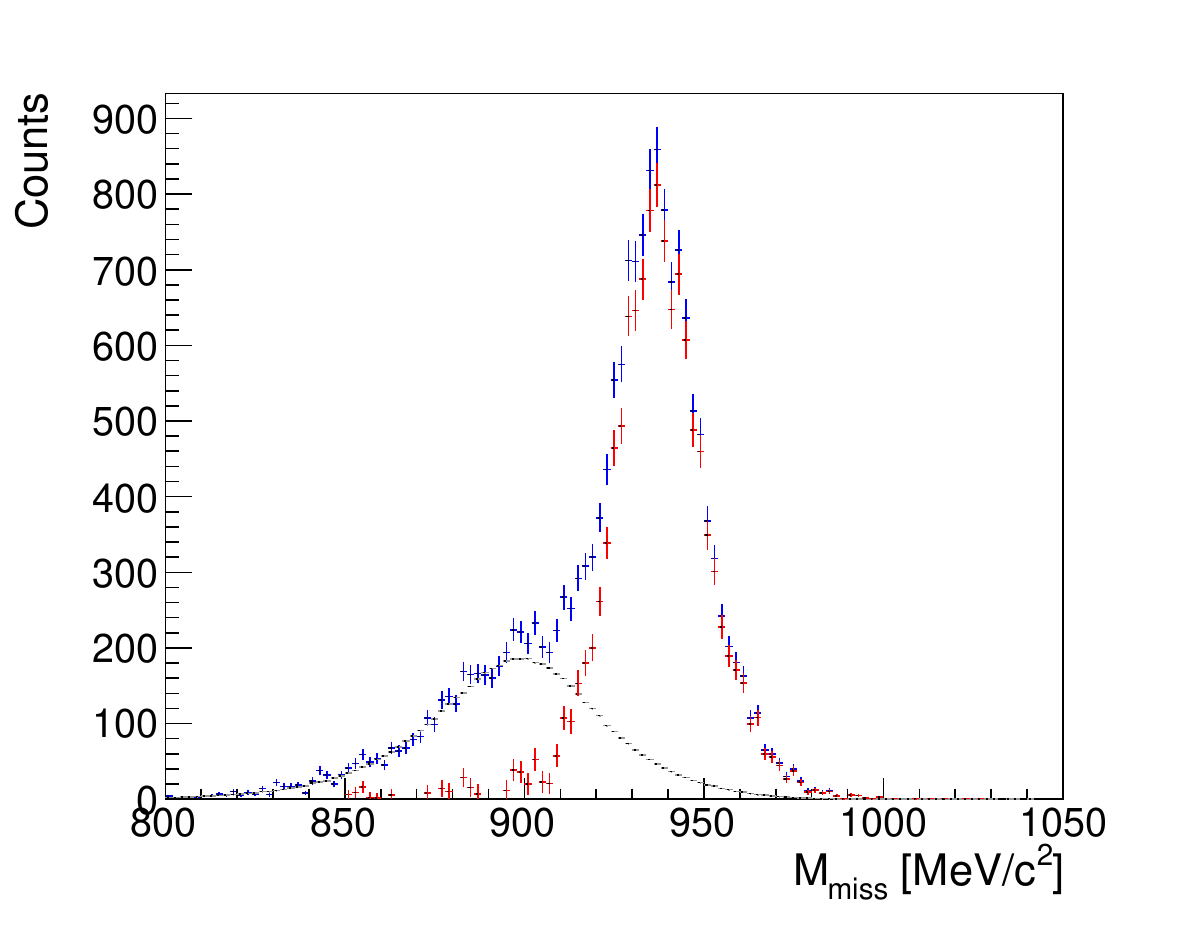}
\caption{\label{fig:time} Left: Time difference $\Delta t$ between the reconstructed $\pi^{0}$ in CB/TAPS (average time of both the decay photons from $\pi^{0} \to 2\gamma$) and the hits in the tagger. Right: sample of the missing mass distribution calculated using eq.~\eqref{eq:missingmass} is shown for the prompt ($\Delta t = 0 \, \pm$~2~ns) and the background sample in blue and gray, respectively. The sample retained after random background subtraction is shown in red.}
\end{figure}

The left panel of figure~\ref{fig:time} shows a sample of the time difference $\Delta t$ distribution between the reconstructed $\pi^0$ in the calorimeters and each hit in the tagger. The so-called prompt peak around $0$~ns corresponding to the coincident hits is well visible. The flat background outside and below the peak instead corresponds to the time-uncorrelated events. The spectrum is generated for the reaction $\gamma p \to p \pi^0$ at a photon beam energy of 240 -- 260~MeV after the application of the kinematic cuts described in subsection~\ref{subsec: input}, used to reduce the initial background contamination of the data.

The prompt time-correlated events are selected in the peak region within an interval $\Delta t \in [-2,2]$~ns (highlighted in blue in the left panel of figure~\ref{fig:time}). To remove the uncorrelated background in the prompt region, the random contribution is modeled by selecting two time windows, one on the left $\Delta t_{l} \in [-200,-50]$~ns, and one on the right $\Delta t_{r} \in [50,600]$~ns of the peak (both highlighted in gray in the left panel of figure~\ref{fig:time}).
The obtained background sample is normalized according to the width of the selected time windows, and then subtracted from the prompt peak sample.

To illustrate the performance of this method, the missing mass distribution was calculated as
\begin{equation}
     	M_{miss} = \sqrt{(E_{\gamma} + m_p  - E_{\pi^0})^2 - (\vec{p}_{\gamma} - \vec{p}_{\pi^0})^2 },
 	\label{eq:missingmass}
\end{equation}
where $\mathbf{p}_{\gamma} = (E_{\gamma}, \vec{p}_{\gamma})$ and $\mathbf{p}_{\pi^0} = (E_{\pi^0}, \vec{p}_{\pi^0})$ are the incoming photon and produced $\pi^0$ four-momenta, respectively, and $m_p$ is the target proton mass at rest.
For a $\gamma p \to p \pi^0$ event, $M_{miss}$ is expected to be in agreement with the proton mass. The obtained distribution is shown in the right panel of figure~\ref{fig:time}, for the prompt and the random sample in blue and gray, respectively, while the distribution obtained after subtracting the uncorrelated events is shown in red. As expected, the shoulder on the left of the total distribution (blue) is well described by the uncorrelated background, resulting in a final distribution well peaked around the proton mass after random background subtraction.

Although this method has been widely used with reliable performance at photon tagging facilities for the past few decades, the need to subtract the random background limits the achievable measurement accuracy, especially for low-energy measurements (where the time uncorrelated background is higher due to the shape of the bremsstrahlung energy distribution) of processes with small cross sections. In addition, the information about all correlations between variables for a single event is lost in the subtraction process (unless the subtraction is performed in multiple dimensions). Thus, a new method allowing for unambiguous matching of the time-correlated tagger hit with the event in the calorimeter without the need for sampling and subtraction of the time-uncorrelated background is highly desirable, especially now that we are entering the so-called precision era of nuclear and hadron physics. This paper presents a new ML-based multivariate analysis method, applicable for the selection of time-correlated hits without subtraction of the time-uncorrelated background.

\section{New analysis concept}
\label{label: concept}

The goal of this work is to distinguish between signal and background events in the region of the prompt peak, shown in the left panel of figure~\ref{fig:time}.
In the newly developed method, ML-based models are created and trained using a simulated data sample with a signature of the reaction of interest in combination with the experimentally measured uncorrelated background\footnote{The simulation of the random background is extremely challenging due to multiple non-linear effects in the experiment.}. The prerequisite for using this method is that the sample of the (potentially) time-correlated events shows patterns similar to the simulated sample. This can be verified, e.g., by comparison of the simulated distributions with the distributions obtained from the experimental data after conventional random background subtraction described in section~\ref{sec:setup}. At the same time, the patterns in the simulated data should be significantly different from the patterns in the random background sample. The selection of input variables (input features for ML models) meeting these criteria allows us to create ML models, trained to distinguish between signal and background events. The performance of the created ML models is tested at first on a data sample consisting of simulated signal events and experimentally measured random background by comparing the initial and predicted labels for both classes of events. After the initial tests, the created ML model is used for the separation of experimentally measured signal and background events in the region of the prompt peak of the time spectrum (discussed in section~\ref{sec:setup} and shown in figure~\ref{fig:time}). Finally, the performance of the ML-based method is compared with the conventional random background subtraction method. 
 
\section{Application of the Machine Learning-based method} 
\label{label: method}

This section illustrates the application of the developed method for the reaction $\gamma p \to p \pi^{0}$ measured with the Crystal Ball/TAPS setup at MAMI for the incoming photon energy range of 240 -- 260~MeV. The simulated Monte Carlo sample, generated with the GEANT4-based~\cite{bib:Geant4} package A2Geant4~\cite{bib:A2Geant4} used to create ML models, consists of $10^6$ $\gamma p \to p \pi^{0}$ events at $E_{\gamma} = 240 - 260$~MeV\footnote{At these energies, there is a significant amount of time-uncorrelated background remaining after the application of routinely used kinematic cuts such as the invariant mass cut, the coplanarity cut, and the opening angle cut.}. The second component used to create ML-based models is the experimentally measured random background sample, described in section~\ref{sec:setup}.

\subsection{Preparation of the input data}
\label{subsec: input}

\begin{figure}[htbp!]
\centering 
\includegraphics[width=\textwidth]{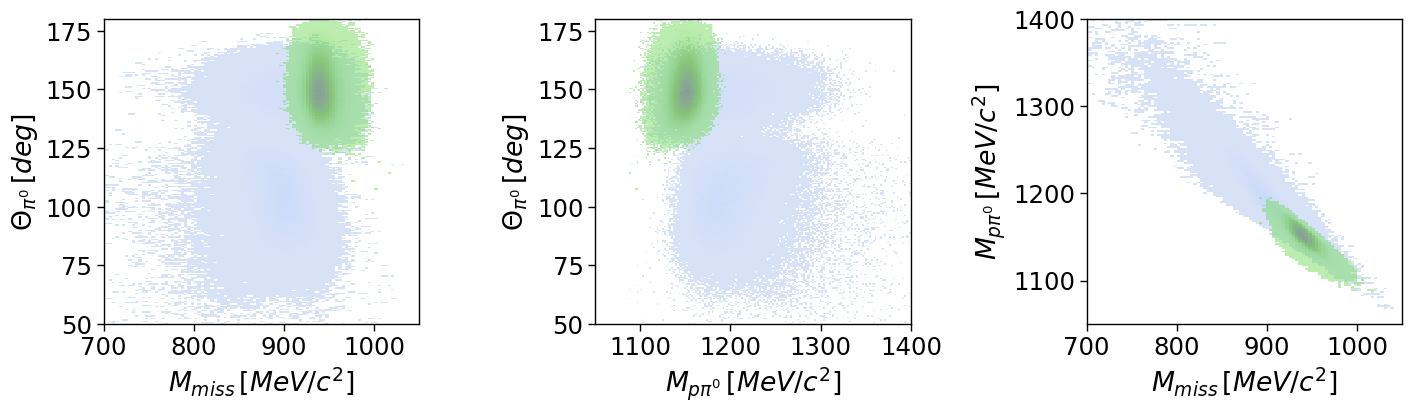}
\includegraphics[width=\textwidth]{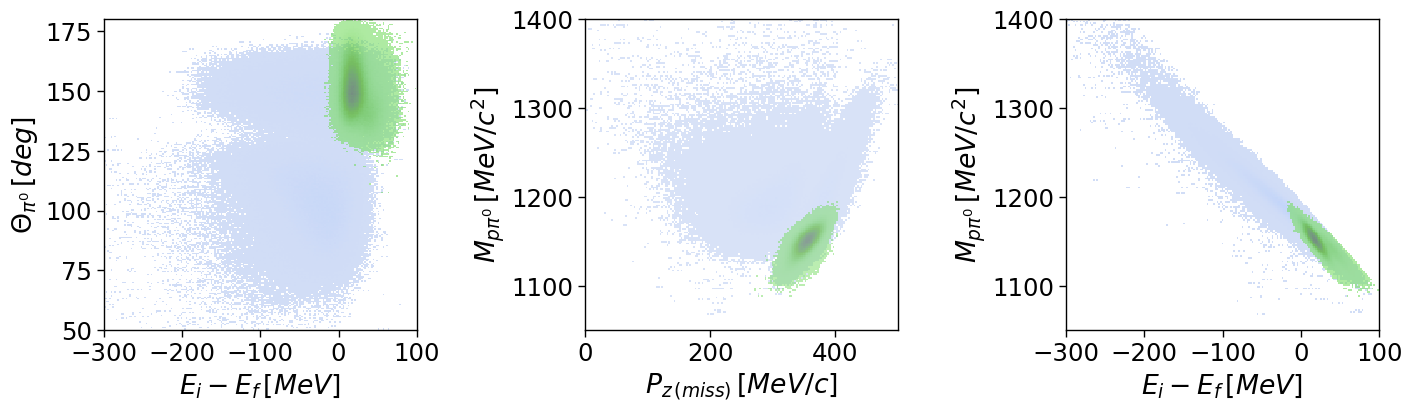}
\caption{\label{fig: initial vars1} Selected correlations of variables for the simulated $\gamma p \to p \pi^{0}$ events (green) and experimentally measured time-uncorrelated background (gray) at $E_{\gamma} = 240 - 260$~MeV. The description of the shown variables is given in the text.}
\end{figure}

To select the $\gamma p \to p \pi^{0}$ reaction, the following cuts were applied both to  the simulated signal and the measured random background. At first, the events with two neutral particles (photon candidates) and one charged particle (proton candidate) were retained. The invariant mass of the two neutral particles had to agree with the nominal $\pi^{0}$ mass within 15~MeV. The difference between the azimuthal angles of the charged particle (proton candidate) and the $\pi^{0}$ (reconstructed from the neutral particles) had to fulfill the condition $|\phi_{p} - \phi_{\pi0}| = 180^{\circ} \pm 10^{\circ}$. The measured polar angle of the proton candidate was matched to the polar angle of the "missing particle" (calculated from the photons in the final and initial states) within $5^{\circ}$. Under these conditions, the events with the signature of the reaction $\gamma p \to p \pi^{0}$ can be clearly identified.

After the event selection described above, an ML model was built using five selected input variables (features). Figure~\ref{fig: initial vars1} shows the comparison between simulated $\gamma p \to p \pi^{0}$ sample (green) and experimentally measured time-uncorrelated background (gray) at $E_{\gamma} = 240 - 260$~MeV. These signal and background samples were used to build an ML-based model trained to distinguish between these two classes of events. In addition to the missing mass, the input variables (features) are the polar angle of $\pi^{0}$ ($\theta_{\pi^0}$), the z-component of the missing momentum ($P_{z (miss)}$), the difference between the energy sum measured for the initial and final states ($E_{i} - E_{f}$), and the invariant mass of the $p\pi^{0}$ pair ($M_{p\pi^{0}}$). 

\subsection{Application of ensemble learning with boosted decision trees using CatBoost}
\label{label: BDT}

The ML algorithm used in this work relies on ensemble learning with gradient boosting for decision trees, where the errors are reduced at each learning step based on the previous step. Generally, gradient boosting algorithms are well suited for solving classification and regression tasks with tabular data. We used the package CatBoost, which is one of the state-of-the-art gradient boosting algorithms, and utilizes symmetric decision trees~\cite{bib:catboost,bib:catboost:arx}. The CatBoost-based models were created using the data sample shown in figure~\ref{fig: initial vars1} as an input.

\begin{figure}[htbp]
\centering 
\includegraphics[width=\textwidth]{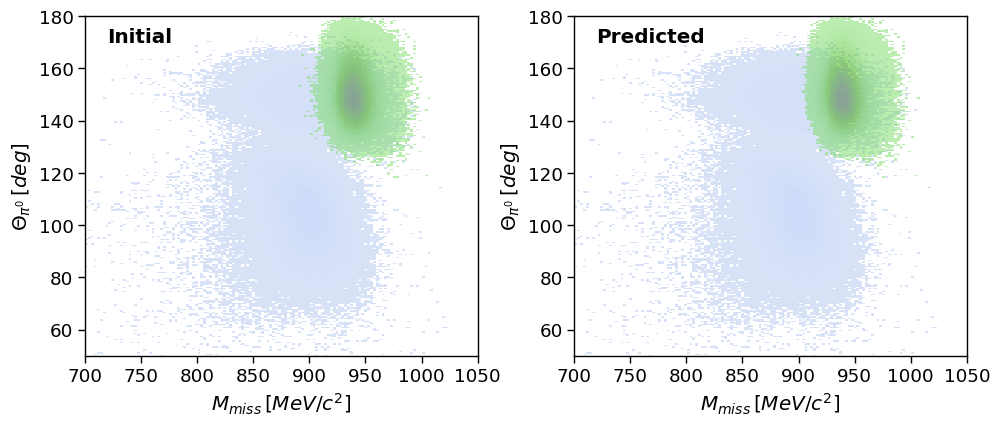}\\
\includegraphics[width=\textwidth]{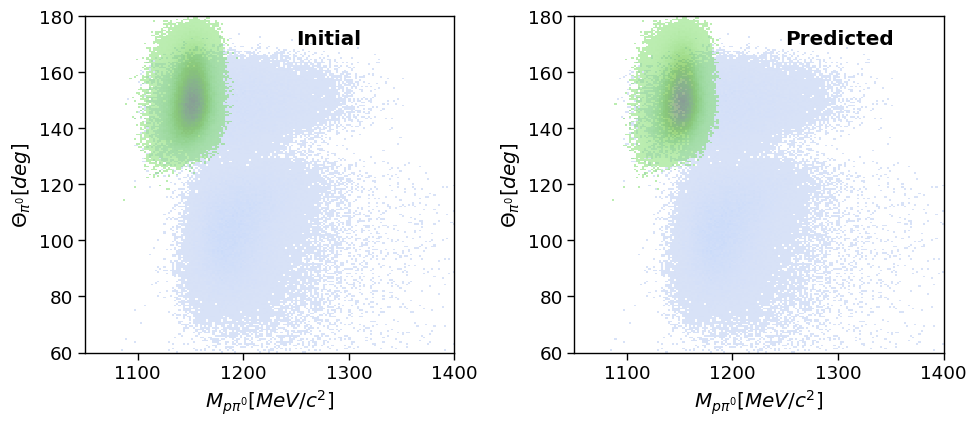}\\
\caption{\label{fig: CatBoost predictions} Comparison of the initially simulated signal events (green) and experimentally measured background (gray) distributions (left panels) and distributions predicted by the trained CatBoost-based model (right panels) for the test data set. The ratio of the number of the simulated signal events and measured uncorrelated events is chosen according to the ratio determined for the data at $\Delta t = 0 \, \pm \, 2 \, \rm ns$ in the prompt peak region.}
\end{figure}

To introduce a realistic ratio for the number of correlated and uncorrelated events in the region of the prompt peak, the ratio of simulated events to random background was chosen to match the ratio of random background to signal events in the prompt peak region (determined from the spectrum shown in figure~\ref{fig:time}). The simulated sample and the measured random background were mixed according to this ratio. Then, the data were randomly reshuffled and split in two parts. The first part containing 2/3 of the data was used to train the ML models, while the remaining 1/3 was used to test the model performance. The models showed high performance with the default settings provided by CatBoost, and were additionally improved by tuning the hyperparameters of the model \footnote{The following hyperparameters were tuned with the so-called Randomized Search method: number of iterations, learning rate, bagging temperature, random strength, and L2 regularization (for details on hyperparameters see ref.~\cite{bib:catboost}).}.
The created models were at first used to separate the events simulated with GEANT4 from the random background. 

Figure~\ref{fig: CatBoost predictions} shows the correlations of the polar angle of the $\pi^0$ vs.\@ the missing mass (top) and the invariant mass of the $p\pi^0$ pair (bottom) for both the initial training sample (left panels) and for the prediction by the CatBoost-based model (right panels). The predicted distributions were obtained using CatBoost-based binary classification, by analyzing the data event-by-event for both the simulated signal and the experimentally measured background events, and then plotting the resulting distribution in the corresponding 2-dimensional histograms.

Part of the background can be linearly separated, however a significant overlap between the two data sets is also present. The shapes of the time-correlated signal (green area in figure~\ref{fig: CatBoost predictions}) are generally well-reproduced by the model. The corresponding values for the precision and recall both for simulated sample and measured random background are summarized in table~\ref{Table: prec_recall} for different prompt peak cuts, corresponding to different levels of background contamination. The precision for the signal events varies between 97.5\%, for the ratio of signal and background events at $\approx 48\%$, and 95\% when the amount of the background increases up to 111.5\% (compared to the number of signal events). As described above, the expected ratios for the signal and background samples were determined for the prompt peak region of the time spectrum (see figure~\ref{fig:time}). In this case, the cut of $0 \, \pm$~2~ns allowed the inclusion of most of the events in the prompt peak, while the broader cuts were used to test the performance of the models in presence of different amounts of background\footnote{For each of the time cuts, a separate model was created, in order to take the ratios of background and signal into account in each case.}. 

\begin{table}[hbtp]
\begin{tabular}{|p{17mm}|p{23mm}|p{20mm}|p{20mm}|p{21mm}|p{21mm}|}
\hline
{\it Time Cut}  & {\it Background/ signal}  & {\it Precision (signal)} & {\it Recall (signal)} & {\it Precision (background)}  &  {\it Recall (background)}\\
\hline
$0 \, \pm$~2~ns & 48.11\% & 0.9750 & 0.9972  & 0.9938 & 0.9467\\
\hline
$0 \, \pm$~3~ns & 67.30\% & 0.9659 & 0.9962  & 0.9940 & 0.9479\\
\hline
$0 \, \pm$~4~ns & 89.25\% & 0.9598 & 0.9931  & 0.9919 & 0.9535\\
\hline
$0 \, \pm$~5~ns & 111.54\% & 0.9506 & 0.9919  & 0.9924 & 0.9537\\
\hline
\end{tabular}
\caption[] {Performance of the CatBoost-based model, trained to separate
the simulated signal and measured background samples. The column {\it Background/signal} represents the ratio of signal and background events determined from the time spectrum shown in figure~\ref{fig:time}.}
\label{Table: prec_recall}
\end{table}

\begin{figure}[htbp]
\centering
\includegraphics[width=\textwidth]{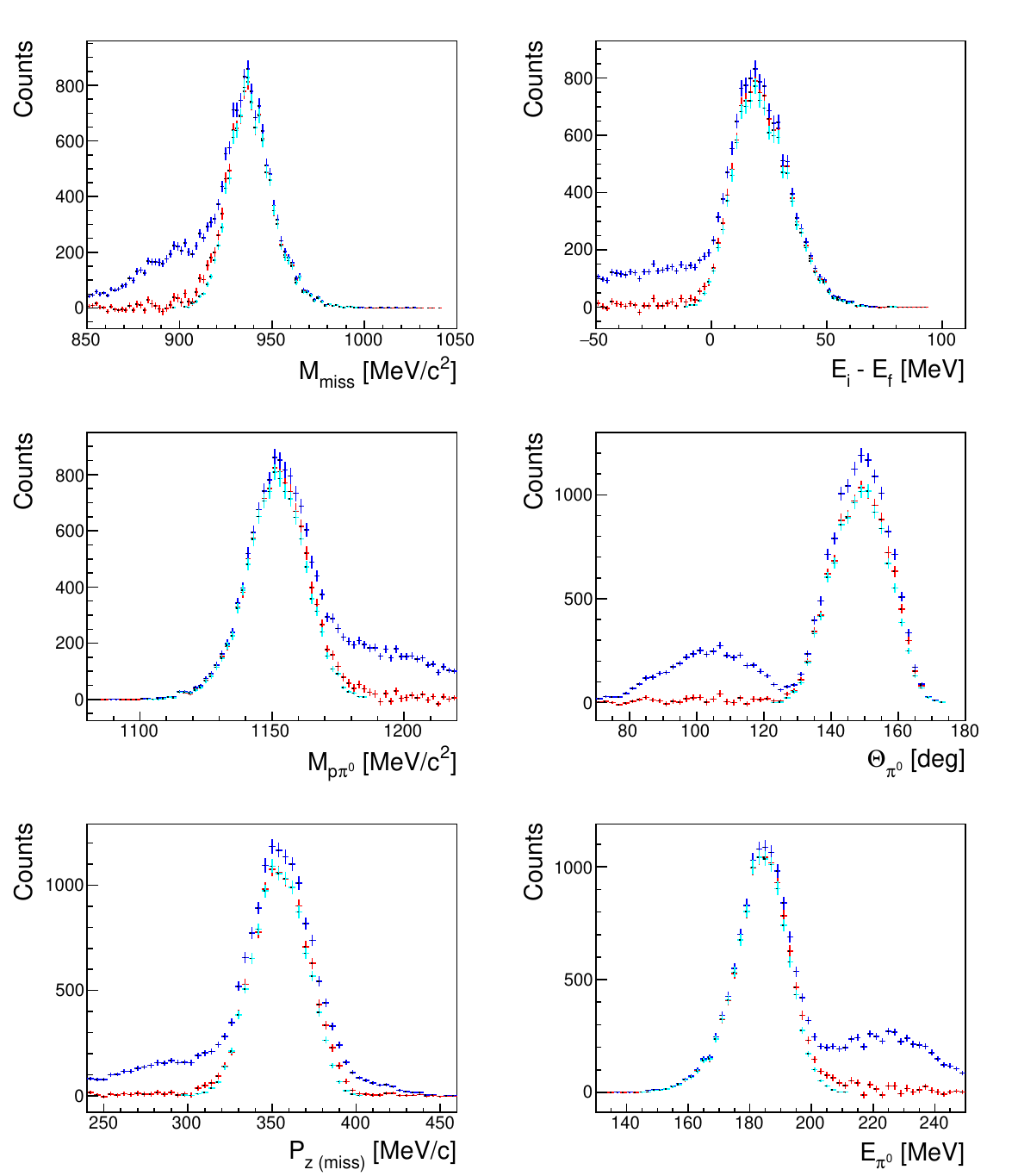}
\caption{\label{fig: CatBoost 6 vars} Comparison of the performance of the ML-based and conventional methods for $\Delta t = 0 \, \pm$~2~ns for all events in the prompt peak region before background subtraction (blue), after background subtraction with application of conventional analysis (red), and predicted with a new ML-based method (cyan).}  
\end{figure}

Finally, the trained and evaluated model was used to distinguish between the time-correlated signal and uncorrelated background in the prompt peak (see figure~\ref{fig:time}). The performance of the CatBoost-based model is compared with the conventional random background subtraction method in figure~\ref{fig: CatBoost 6 vars} for the five variables used as input features for the model (shown in figure~\ref{fig: initial vars1}) and for the total energy of the neutral pion $E_{\pi^{0}}$, not used as an input for the model. The new and the conventional background handling methods show consistent results for most of the bins. The differences observed in the region corresponding to low missing mass values (appearing in the other variables as well), are mainly due to the remaining differences between the simulation (used to build the model) and experimental data and can be improved by further fine-tuning of the simulation (dependent on the goals of the analysis). In the missing mass region above $925 \, \rm MeV/c^{2}$,  the integrals of the distributions obtained with both methods agree on a sub-percent level. 

\begin{figure}    
  \centering
  \begin{tikzpicture}
  \node(picA){\includegraphics[width=\textwidth]{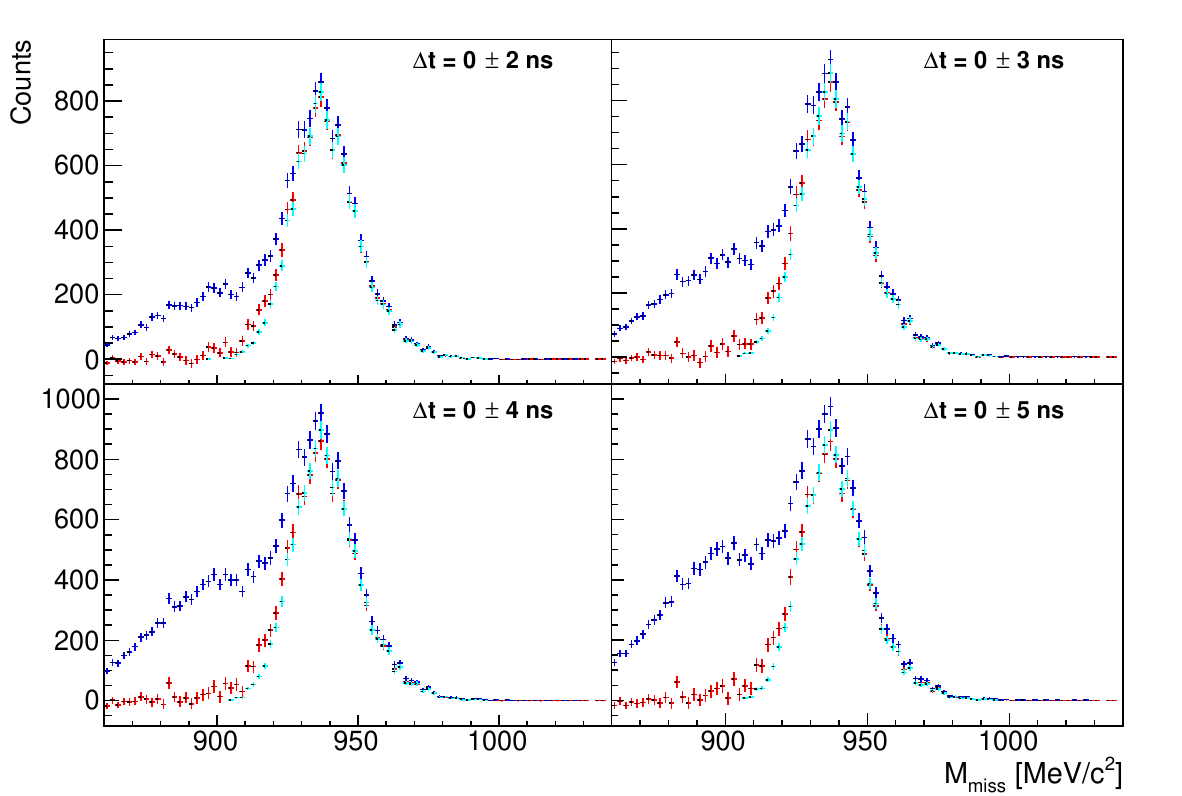}};
  \node at (-1.27, 2.8){\includegraphics[width=0.2\textwidth]{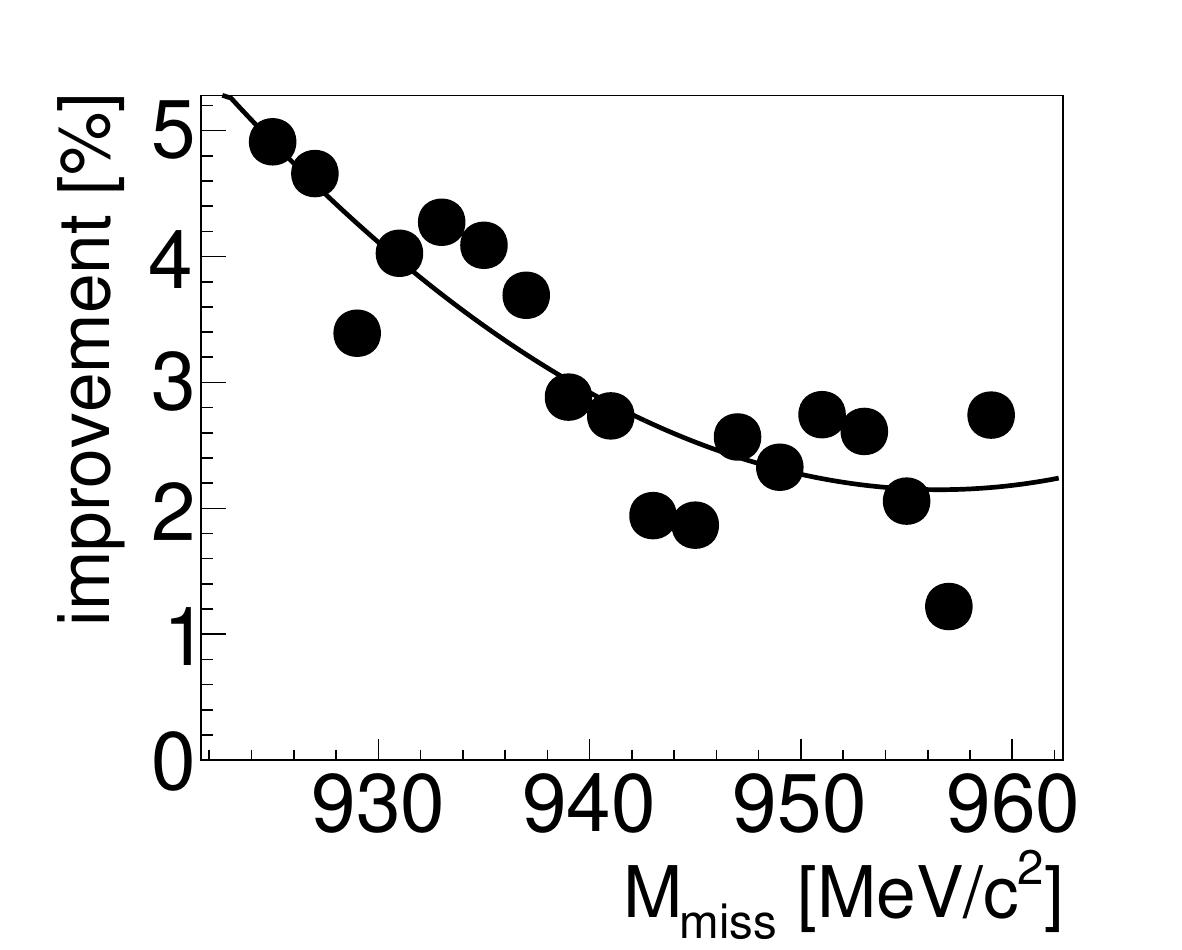}};
    \node at (-1.27, -1.5){\includegraphics[width=0.2\textwidth]{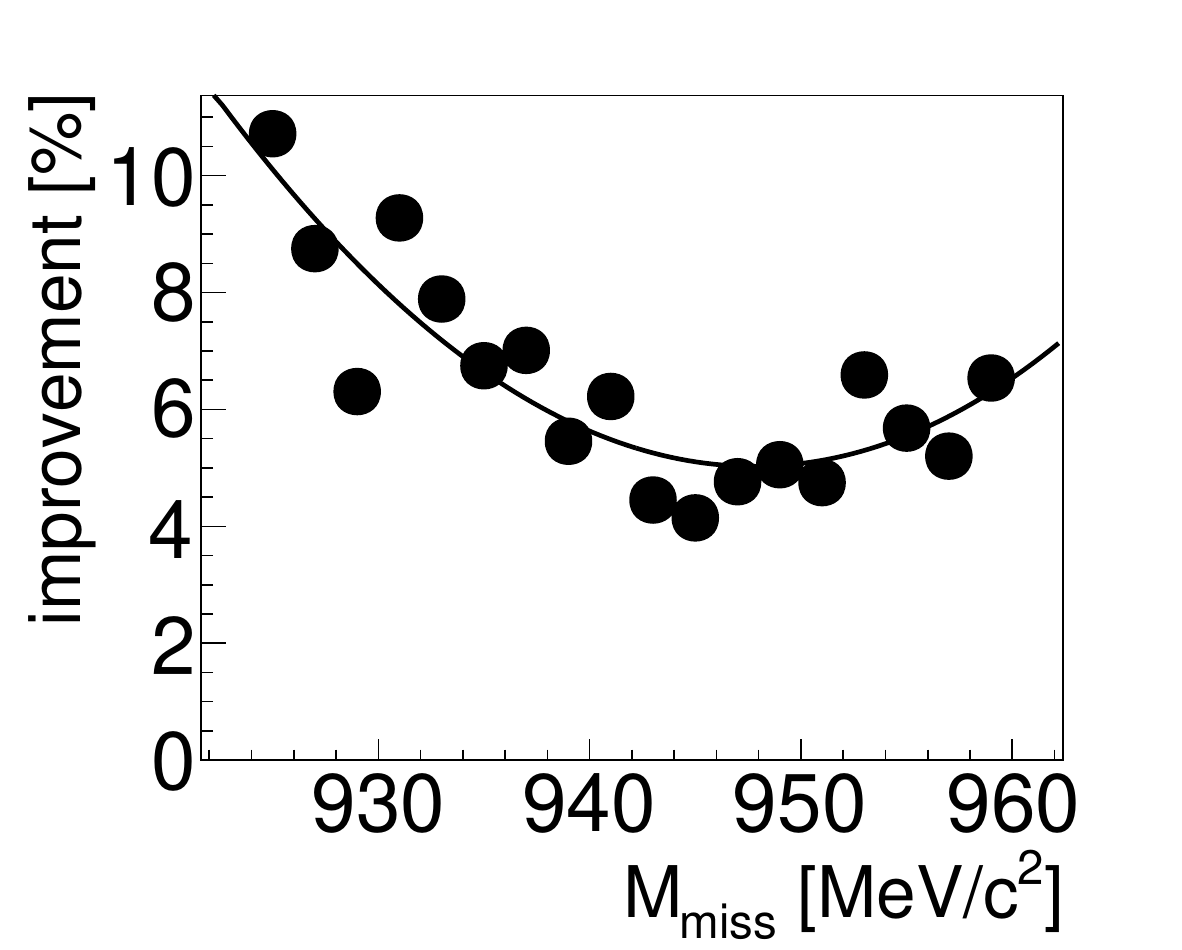}};
  \node at (5.27, 2.8){\includegraphics[width=0.2\textwidth]{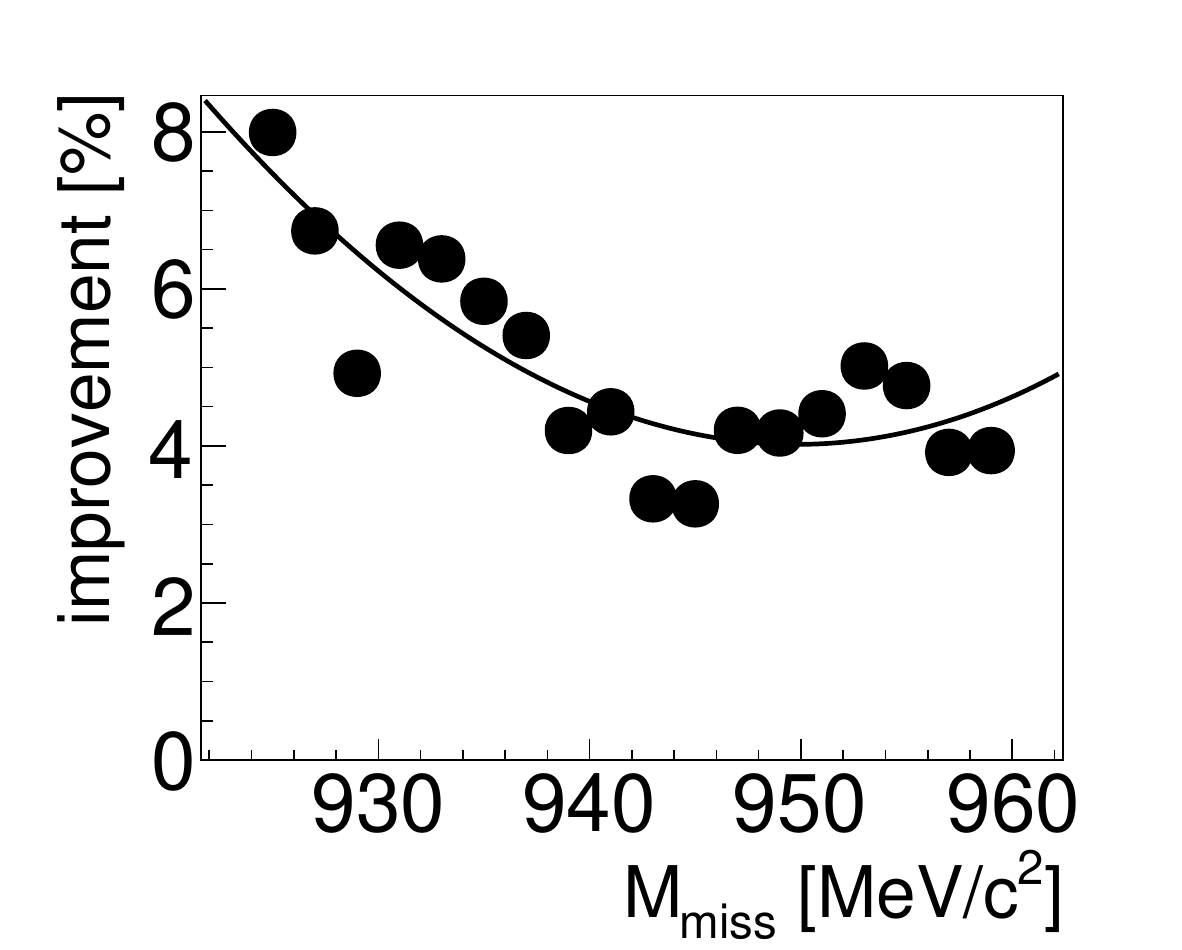}};
  \node at (5.27, -1.5){\includegraphics[width=0.2\textwidth]{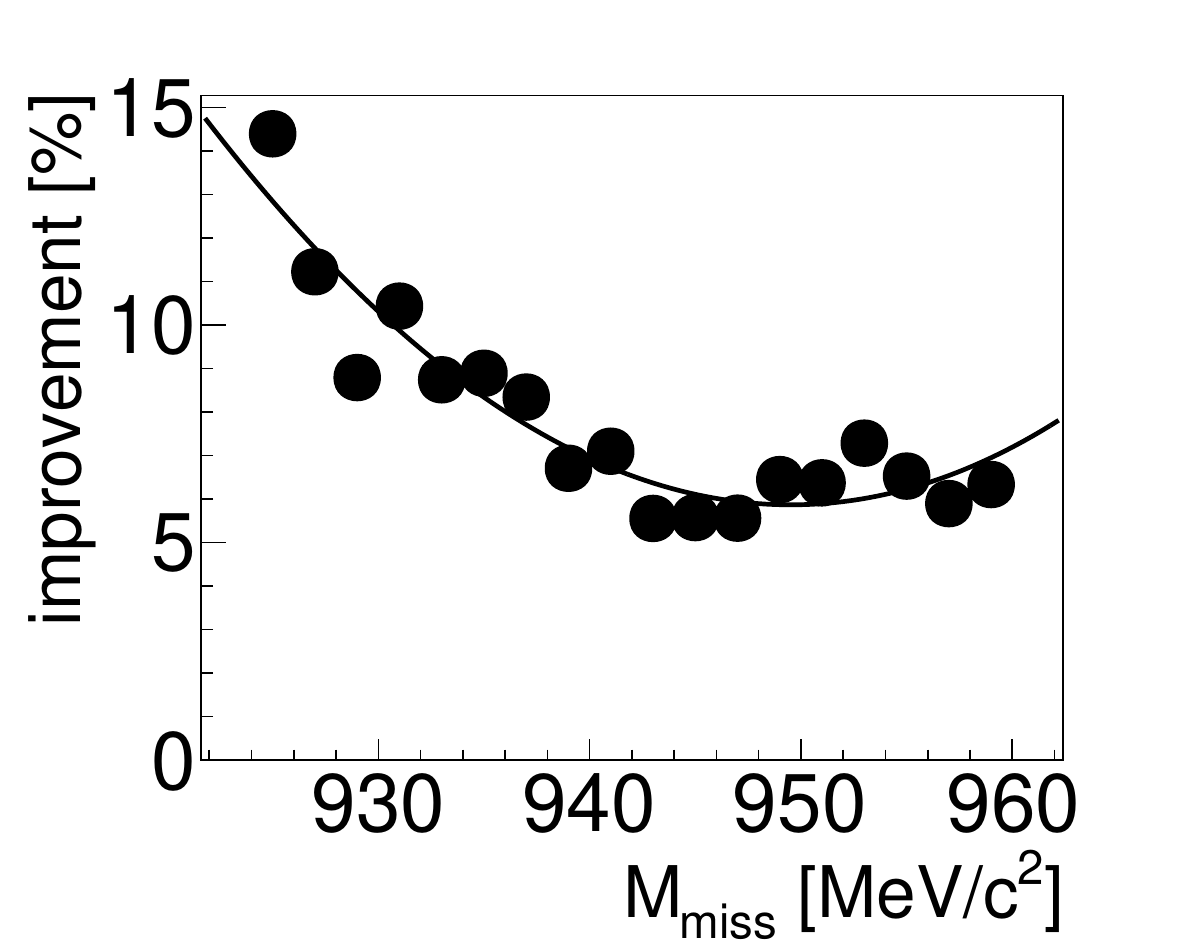}};
    
  \end{tikzpicture}
  \caption{\label{fig: CatBoost DeltaT} Missing mass distributions for all events in the prompt peak region (blue), after background subtraction with application of conventional analysis (red), and predicted with a new ML-based method (cyan) for different $\Delta t$ cuts: $0 \, \pm$~2~ns, $0 \, \pm$~3~ns, $0 \, \pm$~4~ns, and $0 \, \pm$~5~ns. The insets show the improvement in the relative uncertainties (in percent) due to the application of the new ML-based method with respect to the standard background subtraction method (the curves representing a fit with a third-order polynomial are plotted to guide the eye).}
\end{figure}

It is important to note that the developed method is not only applicable for the variables used as input features to the model, but can also be used to predict distributions of other variables, which are correlated with the input features used to build the models, as can be seen 
for the total energy of the neutral pion $E_{\pi^{0}}$ in the lower right panel of figure~\ref{fig: CatBoost 6 vars}. Moreover, since the new method does not require subtraction of the random background, the resulting uncertainties are smaller compared to the uncertainties for the conventional method, even though the magnitude of the reduction depends strongly on the amount of background in the corresponding analysis.

In addition, the performance of the developed method and conventional background subtraction were compared at different levels of background contamination.
Figure~\ref{fig: CatBoost DeltaT} shows the comparison between missing mass spectra corresponding to different prompt peak cut widths, resulting in different background contamination of the data. For each of these cases, corresponding to different prompt peak cuts, CatBoost-based models were built separately, taking into account the expected ratio of the correlated signal to uncorrelated background (dependent on the width of the selected time window)\footnote{The tests with using the same ratio for different cases showed rather stable results (in particular for the cases with not very different ratios), however, the best agreement with the well-established subtraction method was achieved by adjusting the ratios separately for each case.}. Generally, the comparison of the results obtained with the new ML-based approach with the conventional random background subtraction indicates stable model performances at significantly different levels of random background.
The differences at low missing mass values, as mentioned above, can be additionally suppressed (dependent on the goals of the analysis) by further reduction of differences between GEANT simulation and experimental data (used to build the ML-based models). The integrals shown in table~\ref{Table: bg levels} for the new and conventional analysis methods are in agreement within $\approx 1\% $ for the missing mass range of  $925 - 960 \, \rm MeV/c^{2}$.

The insets in figure 6 show the improvement in the resulting uncertainties after applying the new ML-based method compared to the standard background subtraction method for the missing mass range of  $925 - 960 \, \rm MeV/c^{2}$ (calculated from the ratio of the relative errors for both methods). The most significant improvement (5\% -- 15\%) is achieved at lower missing masses, where the background level is higher in this given case. It is also worth noting the greater improvement at wider time cuts, since the overall background level depends on the width of the time cuts.

\begin{table}[hbtp]
\centering
\begin{tabular}{|p{20mm}|p{30mm}|p{25mm}|}
\hline
{\it Time Cut} & {\it Background/signal}  & {\it $N_{sub}/N_{ML}$} \\

\hline
$0 \, \pm$~2~ns & 48.11\% & 0.9977\\
\hline
$0 \, \pm$~3~ns & 67.30\% & 0.9948 \\
\hline
$0 \, \pm$~4~ns & 89.25\% & 0.9936\\
\hline
$0 \, \pm$~5~ns & 111.54\% & 0.9920\\
\hline
\end{tabular}
\caption[] {Comparison of the performance of the ML-based model and conventional background subtraction. $N_{ML}/N_{sub}$ indicates the ratio of the integrals
of the missing mass spectra at $925 - 960 \, \rm MeV/c^{2}$.}
\label{Table: bg levels}
\end{table}

Finally, it is worth noting that the new ML-based method allows us to preserve the correlation of all variables in the event-by-event analysis of experimental data. In the commonly used standard method of background subtraction, e.g., for one- or two-dimensional histograms (without considering all variables describing the studied reaction), the information about individual events, including the correlations of the variables describing each event, can be lost. The method proposed in this work allows the separation of time-correlated and uncorrelated events on an event-by-event basis, without the need to sample and subtract the corresponding distributions (histograms), thus preserving the correlations between all variables for each individual event. Therefore, the data sets obtained with this method contain more information than those extracted with direct background sampling and subtraction. In many cases, this information can be particularly valuable for further phenomenological analyses of the experimental results.

In analogy to the data set obtained with the setup of the A2 Collaboration at MAMI and analyzed with the presented ML-based method, the measurement uncertainties in other experiments using tagged photons at high rates can be reduced, especially when the final uncertainties are significantly limited by the presence of a significant amount of time-uncorrelated background.

\section{Summary}
\label{label: Summary}

A newly developed Machine Learning-based method for the selection of the time-correlated signal at tagged photon facilities is presented. The application of this method allows to reduce the uncertainties of the measurements at experiments, where the conventional sampling and subtraction of the uncorrelated background poses restrictions on the accuracy of the measurements. Moreover, the developed method allows to analyze data event-by-event, thus preserving the information about the correlations of the variables for individual events, in contrast to the standard subtraction method. The new method shows stable performance in handling data with different levels of background contamination. One of the future applications of this method will be the analysis of the Compton scattering data taken with hydrogen and light nuclear targets in order to improve the accuracy of the extraction of the scalar and spin polarizabilities of the nucleons. 
 
The scripts used in this work as well as parts of the data used in this
study are available upon request.

\acknowledgments
The data used in this work were taken by the A2 Collaboration with
the Crystal Ball/TAPS setup at MAMI in March 2018.
E.~M.\@ acknowledges the support by the Deutsche Forschungsgemeinschaft
(DFG, German Research Foundation) through the Research Unit [Photon-photon interactions in the Standard Model and beyond, Projektnummer 458854507 - FOR 5327].

\end{document}